\documentclass[twocolumn]{aastex62}
\usepackage[utf8]{inputenc}
\DeclareUnicodeCharacter{2212}{-}
\usepackage{graphicx}
\graphicspath{{./}{Figures}}
\usepackage{amsmath}
\usepackage{afterpage}
\usepackage{enumitem}
\usepackage{xcolor}
\usepackage{float}
\usepackage{xspace} 
\usepackage{url}
\usepackage{fontawesome}
\setenumerate{itemsep=0mm}
\usepackage{wrapfig}
\usepackage{placeins}
\newcommand{\uno}{Ursa~Major~III/UNIONS~1}
\newcommand{\unoshort}{UMaIII/U1}
\defcitealias{2024ApJ...961...92S}{S24}
\newcommand\kms{\mbox{km\,s$^{\rm -1}$}}

\usepackage{amsmath}
\usepackage{amssymb}
\usepackage{xspace}
\usepackage{xifthen}
\usepackage{eso-pic}

\newcommand{\Gaia}{{\it Gaia}\xspace}

\definecolor{forestgreen}{HTML}{228B22}
\definecolor{urlblue}{HTML}{000000}

\mathchardef\mhyphen="2D

\newlength{\dhatheight}

\newcommand{\code}[1]{\texttt{#1}\xspace}

\newcommand{\degree}{\ensuremath{{}^{\circ}}\xspace}

\newcommand{\secref}[1]{Section~\ref{sec:#1}}

\newcommand{\tabref}[1]{Table~\ref{tab:#1}}

\newcommand{\figref}[1]{Figure~\ref{fig:#1}}

\newcommand{\bandvar}[2][]{%
  \ifthenelse{\isempty{#1}}{\var{#2}}{\var{#2\_#1}}%
}

\newcommand{\var}[1]{\ensuremath{\texttt{\MakeUppercase{#1}}}\xspace}

\providecommand\physrep{\ref@jnl{Phys.~Rep.}}%
\providecommand\apjs{\ref@jnl{ApJS}}%
\providecommand{\jcap}{\ref@jnl{JCAP}}%

\newcommand{\vdispFreqCompleteEtwo}{2.3}
\newcommand{\vdispFreqPureEtwo}{1.6}
\newcommand{\WolfMassFreqCompleteEtwo}{10670}
\newcommand{\WolfMassFreqPureEtwo}{4970}
\newcommand{\MLRatioFreqCompleteEtwo}{1940}
\newcommand{\MLRatioFreqPureEtwo}{900}
\newcommand{\vdispBayesCompleteEtwo}{2.6}
\newcommand{\vdispBayesPureEtwo}{2.2}

\newcommand{\BayesFactorCompleteEtwo}{-3.7}
\newcommand{\BayesFactorPureEtwo}{-4.0}
\newcommand{\vdispBayesLogCompleteEtwo}{1.7}
\newcommand{\vdispBayesLogPureEtwo}{1.4}

\newcommand{\WolfMassBayesLogPureEtwo}{3940}

\newcommand{\MLRatioBayesLogPureEtwo}{720}

\newcommand{\JfactorFreq}{20.2}

\newcommand{\BayesianFeH}{$-2.65~\pm~0.1$}
\newcommand{\BayesianFeHDispUL}{0.35}
\newcommand{\BayesFactorFeH}{-3.1}

\shorttitle{A New Observational Paradigm for \unoshort }

\shortauthors{Cerny, Bissonette, and Ji et al.}

\begin{document}

\reportnum{\footnotesize}

\title{No Observational Evidence for Dark Matter Nor a Large Metallicity Spread \\ in the Extreme Milky Way Satellite Ursa Major III / UNIONS~1}

\correspondingauthor{William Cerny}
\email{william.cerny@yale.edu}

\author[0000-0003-1697-7062]{William~Cerny}
\affiliation{Department of Astronomy, Yale University, New Haven, CT 06520, USA}

\author[0009-0006-5977-618X]{Daisy~Bissonette}
\affiliation{Department of Astronomy and Astrophysics, University of Chicago, Chicago, IL 60637, USA}

\author[0000-0002-4863-8842]{Alexander~P.~Ji}
\affiliation{Department of Astronomy and Astrophysics, University of Chicago, Chicago, IL 60637, USA}
\affiliation{Kavli Institute for Cosmological Physics, University of Chicago, Chicago, IL 60637, USA}
\affiliation{NSF-Simons AI Institute for the Sky (SkAI), 172 E. Chestnut St., Chicago, IL 60611, USA}

\author[0000-0002-7007-9725]{Marla~Geha}
\affiliation{Department of Astronomy, Yale University, New Haven, CT 06520, USA}

\author[0000-0002-7155-679X]{Anirudh~Chiti}
\affiliation{Department of Astronomy and Astrophysics, University of Chicago, Chicago, IL 60637, USA}
\affiliation{Kavli Institute for Cosmological Physics, University of Chicago, Chicago, IL 60637, USA}

\author[0000-0002-6946-8280]{Simon~E.T.~Smith}
\affiliation{Department of Physics and Astronomy, University of Victoria, Victoria, BC, V8P 1A1, Canada}

\author[0000-0002-4733-4994]{Joshua~D.~Simon}
\affiliation{Observatories of the Carnegie Institution for Science, 813 Santa Barbara St., Pasadena, CA 91101, USA}

 \author[0000-0002-6021-8760]{Andrew~B.~Pace}
\affiliation{Department of Astronomy, University of Virginia, 530 McCormick Road, Charlottesville, VA 22904 USA}

\author[0000-0002-4733-4994]{Evan~N.~Kirby}
\affiliation{Department of Physics and Astronomy, University of Notre Dame, 225 Nieuwland Science Hall, Notre Dame, IN 46556, USAA}		

\author[0000-0003-4134-2042]{Kim A. Venn}
\affiliation{Department of Physics and Astronomy, University of Victoria, Victoria, BC, V8P 1A1, Canada}

\author[0000-0002-9110-6163]{Ting~S.~Li}
\affiliation{Department of Astronomy and Astrophysics, University of Toronto, 50 St. George Street, Toronto ON, M5S 3H4, Canada}

\author[0009-0009-9570-0715]{Alice~M.~Luna}
\affiliation{Department of Astronomy and Astrophysics, University of Chicago, Chicago, IL 60637, USA}
\affiliation{Kavli Institute for Cosmological Physics, University of Chicago, Chicago, IL 60637, USA}

\begin{abstract}
The extremely-low-luminosity, compact Milky Way satellite Ursa Major III / UNIONS~1 (UMaIII/U1; $L_V = 11 \ L_{\odot}$; $a_{1/2} = 3$~pc) was found to have a substantial velocity dispersion at the time of its discovery ($\sigma_v = 3.7^{+1.4}_{-1.0} \rm \ km \ s^{-1}$), suggesting that it might be an exceptional, highly dark-matter-dominated dwarf galaxy with very few stars. However, significant questions remained about the system's dark matter content and nature as a dwarf galaxy due to the small member sample ($N=11$), possible spectroscopic binaries, and the lack of any metallicity information. Here, we present new spectroscopic observations covering $N=16$ members that both dynamically and chemically test UMaIII/U1's true nature. From higher-precision Keck/DEIMOS spectra, we find a $95\%$ confidence level velocity dispersion limit of $\sigma_v< 2.3 \rm \ km \ s^{-1}$, with a $\sim$120:1 likelihood ratio favoring the expected stellar-only dispersion of $\sigma_* \approx 0.1 \rm \ km \ s^{-1}$ over the original $3.7 \rm \ km \ s^{-1}$ dispersion. There is now no observational evidence for dark matter in the system. From Keck/LRIS spectra targeting the Calcium II K line, we also measure the first metallicities for 12 member stars, finding a mean metallicity of $\rm [Fe/H] = -2.65 \; \pm \, 0.1$ (stat.) $\pm~0.3$ (zeropoint) with a metallicity dispersion limit of $\sigma_{\rm [Fe/H]} < 0.35$~dex (at the 95\% credible level). Together, these properties are more consistent with UMaIII/U1 being a star cluster, though the dwarf galaxy scenario is not fully ruled out. Under this interpretation, UMaIII/U1 ranks among the faintest and most metal-poor star clusters yet discovered. 
\end{abstract}

\keywords{dwarf galaxies; star  clusters}


\section{Introduction}
\label{sec:intro}
\par \uno{}, discovered by \citealt{2024ApJ...961...92S} (hereafter \citetalias{2024ApJ...961...92S}) in the Ultraviolet Near Infrared Optical Northern Survey (UNIONS; \citealt{2025arXiv250313783G}) is the most extreme Milky Way satellite yet discovered --- and one of the most ambiguous in nature. With a total luminosity of just $L_V = 11 \ L_\odot$, the system is the faintest ancient stellar population known in the universe and $\sim$4-5$\times$ less massive than the next faintest Milky Way satellites (Kim~3 and DELVE~5; \citealt{2016ApJ...820..119K,2023ApJ...953....1C}). Although its small  physical size (elliptical half-light radius of $a_{1/2} = 3$~pc) is far more consistent with known globular clusters than confirmed ultra-faint dwarf galaxies, initial spectroscopic observations of \unoshort{} with the 10-m Keck II telescope and its DEep Imaging Multi-Object Spectrometer (DEIMOS; \citealt{2003SPIE.4841.1657F}) presented by  \citetalias{2024ApJ...961...92S} revealed a substantial velocity dispersion of $\sigma_v = 3.7^{+1.4}_{-1.0} \rm \ km \ s^{-1}$, consistent with the presence of a massive dark matter halo for the system and far above the predicted dispersion of $\sigma_* \sim 0.1 \rm \ km \ s ^{-1}$ expected if the system is comprised solely of stars and stellar remnants \citep{2025MNRAS.539.2485D}. \par The enormous mass-to-light ratio implied by this velocity dispersion, $M/L_V = 6500 \,\rm \,M_\odot L_\odot^{-1}$,  suggested the system might be one of the smallest, densest, and most dark-matter-dominated galaxies yet discovered.  In possible support of this scenario,  \citet{2024ApJ...965...20E} presented dynamical models of \unoshort{} demonstrating that the system should have tidally disrupted   within two orbital periods if it was a self-gravitating cluster --- providing circumstantial evidence for the need for dark matter to stabilize the system against Milky Way tides. If confirmed, these remarkable properties would make the system a  superlative testbed for dark matter physics, especially for efforts to indirectly detect dark matter through its annihilation or decay into Standard Model particles \citep{2024PhRvD.109j3007B,2024PhRvD.109h3018C,2024ChPhC..48k5112Z, 2024PhRvD.110l3026Z,2024PhRvD.110j3048Z,2025PhRvD.111f3078H,2025arXiv250803823F}, tests of the tidal evolution of dark matter subhalos and their inner density profiles \citep{2024ApJ...968...89E,2024ApJ...965...20E}, and constraints on the allowed mass range of ultra-light dark matter particles \citep{2024ApJ...965...20E,2025arXiv250902781M}. This is particularly true because the system is closer than any known galaxy or galaxy candidate ($D_{\odot} = 10  \pm 1$~kpc). 
\par However, as duly noted by \citetalias{2024ApJ...961...92S}, there were immediate reasons to be cautious in interpreting this high dark matter content. The removal of a single star in the sample of 11 members from DEIMOS reduced the measured velocity dispersion by a factor of two, and the removal of a second reduced the resolved dispersion to an upper limit. A clear concern was that one or more member stars could be binary systems observed far from their center-of-mass velocities, biasing our view of \unoshort{}'s stellar kinematics toward the dark matter scenario.  More recently, \citet{2025MNRAS.539.2485D} have also showed that the system's long-term survival could be facilitated by the retention of a substantial population of stellar remnants instead of dark matter, and that even the system's high velocity dispersion could be explained in the stellar-only case if the system exhibited a large primordial binary fraction. \citet{Rostami-Shirazi_2025} similarly demonstrated that the initial observations were consistent with a self-gravitating ``dark star cluster'' scenario in which \unoshort{} harbors a significant black hole population with weak natal kicks. These lines of evidence demonstrate that both the system's dark matter content and its very classification as a suspected dwarf galaxy are far from certain.
\par In this \textit{Letter}, we challenge the initial observational picture of \uno{} as a dark-matter-dominated dwarf galaxy candidate based on new, deep Keck spectroscopy of the system covering an expanded sample of members. Our results demonstrate there is now no positive evidence for dark matter in the system (\secref{dynamics}) nor any evidence for large internal metallicity variations (\secref{chemistries}) typical of dwarf galaxies. We ultimately argue that the system is more likely to be the lowest-mass old star cluster yet discovered  (\secref{classification}).  We summarize our results in \secref{summary}.

\begin{figure*}
    \centering
    \includegraphics[width=0.98\textwidth]{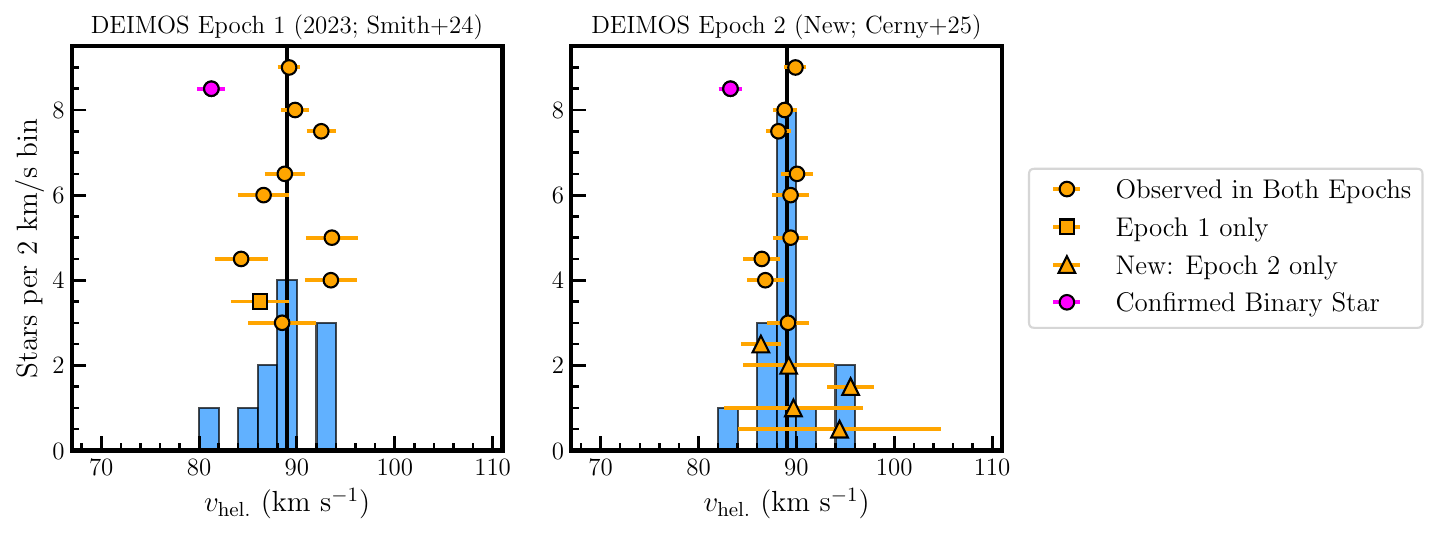}
    \caption{\textbf{The velocity distribution of UMaIII/U1 members} from the re-reduced 2023 epoch (left) and the deeper April 2025 epoch (right), all from  Keck/DEIMOS. The original 2023 epoch identified 11 members; in 2025 we successfully re-observed 10 of these stars with DEIMOS and added five new members (including two with good precision and three with larger uncertainties). In the deeper second epoch, which yielded significantly more precise measurements, we observe a clear regression to the mean in most member stars. The binary star that we monitored (\texttt{S24\_M2}; in magenta) remains a clear outlier in each panel, but we note that the evidence for binarity is based primarily on the GMOS/MagE/HIRES data not shown here.}
    \label{fig:histograms}
\end{figure*}

\section{No Observational Evidence for Dark Matter  in UM\lowercase{a}III/U1}
\label{sec:dynamics}
\subsection{Keck/DEIMOS Observations and Member Sample}
\par To improve upon the dynamical constraints reported by \citetalias{2024ApJ...961...92S} (hereafter, Epoch 1 or the 2023 epoch),  we re-observed \uno{} with Keck/DEIMOS on UTC 2025-04-02. We integrated for 4.66 hours (compared to $1$ hour in 2023) through a new multi-slit mask designed to cover the previously-confirmed \citetalias{2024ApJ...961...92S} members and new faint candidates selected from the UNIONS photometric catalogs. We used the instrument's 1200G grating and OG550 order-blocking filter. For our mask's 0.7\arcsec{} slits, this yielded $R \approx 6000$ across a typical wavelength range $6500 \rm \, \AA$--$9000 \rm \, \AA$. We reduced this data with \texttt{PypeIt} \citep{2020JOSS....5.2308P} and subsequently derived stellar velocities using the \texttt{dmost} pipeline (Geha et al., in press).  The effective per-mask systematic velocity uncertainty floor for this instrument setup and pipeline is $1.1~\kms$.  These procedures are extensively described and validated in Geha et al., in press, so we do not detail them further here. We note that we re-reduced the \citetalias{2024ApJ...961...92S} spectra, incorporating small updates to the velocity uncertainty prescription and a global $0.1~\kms{}$ velocity zeropoint shift. The updated measurements are reported in \tabref{members}.
\par From this second-epoch DEIMOS data alone, we identified a total of 15 velocity members including 10 of the previously-reported Epoch 1 members and 5 new members (see \figref{histograms} and \tabref{members}).  We enumerate the former 10 as \texttt{S24\_M\#}, where \# refers to the associated index given in Table 2 of \citetalias{2024ApJ...961...92S}, and enumerate the latter five  (new) members as \texttt{C25\_M\#}. One clear member star in Epoch 1 (\texttt{S24\_M12}) was not recovered in Epoch 2 because it fell in a chip gap. Thus, the complete member sample across both epochs totals 16 stars. All 16 of these stars form a coherent velocity peak distinct from the empirical foreground distribution, exhibit self-consistent \Gaia DR3 proper motions (\citealt{2023A&A...674A...1G}; where available), and fall along a $\tau = 12$~Gyr, $\rm [Fe/H]_{\rm iso} = -2.19$  isochrone at a distance modulus of $(m-M)_0 = 15.0$ (parameters from \citetalias{2024ApJ...961...92S}, model from PARSEC v1.2S; \citealt{2012MNRAS.427..127B}). The membership for the system therefore remains unambiguous; see Appendix \ref{sec:appAdiagnostic} for a diagnostic figure. It bears emphasizing that the narrow spread in observed velocities confirms that \unoshort{} is a bound stellar system.

\subsection{Confirmation of Star \texttt{S24\_M2} as a Binary}
\label{sec:binconf}
\citetalias{2024ApJ...961...92S} speculated that the member star \texttt{S24\_M2} may be a binary system because of its outlying velocity in the 2023 DEIMOS data. To test this hypothesis, we monitored this star for short-period radial velocity variations across May 2025 through targeted single-object observations. Specifically, we obtained additional spectra of this star with Magellan/MagE \citep{2008SPIE.7014E..54M} on UTC 2025-05-04 (55 minutes, $0.7\arcsec$ slit; $R\approx 6100$), Gemini North/GMOS \citep{2004PASP..116..425H,2016SPIE.9908E..2SG}\footnote{Observed in queue/service mode through Gemini Fast Turnaround program GN-2025A-FT-110; PI: Cerny} on 2025-05-25 (43 minutes, $0.5\arcsec$ slit; $R\approx 4900$), and Keck/HIRES-r \citep{1994SPIE.2198..362V} on 2025-05-28 ($\sim$1.6 hours, D3 decker/$1.7\arcsec$ slit; $R\approx 24000$).  Calibrations were taken after the science exposures in each case to capture instrument flexure at the time of observation. These data were reduced with the \texttt{CarPy} \citep{2000ApJ...531..159K,2003PASP..115..688K},  \texttt{PypeIt} \citep{2020JOSS....5.2308P}, and \texttt{MAKEE}\footnote{\url{https://sites.astro.caltech.edu/~tb/makee/}} pipelines, respectively. Velocities and uncertainties were derived by calculating $\chi^2$ of individual spectral orders against a Doppler-shifted synthetic spectrum (see \citealt{Ji2020a}). Star-slit miscentering was corrected using telluric absorption templates from \texttt{Telfit} \citep{2014AJ....148...53G}. We added systematic error floors of $1.7$~\kms{} and $4$~\kms{} in quadrature for MagE and GMOS, respectively (\citealt{2025arXiv250804053C,2023MNRAS.518.1057E}). For HIRES-r, we assumed a conservative $1$~\kms{} floor but the true precision may be better.  
\par The resulting velocity measurements, provided in \tabref{binary}, demonstrate clear velocity variation across our more recent epochs -- especially between our MagE measurement and the HIRES and GMOS datapoints taken three weeks thereafter. Based on a $\chi^2$ test against the median velocity across epochs, we reject the null hypothesis of no velocity variation at the $p < 0.05$ level ($p = 0.023$). We therefore confirm star \texttt{S24\_M2} as binary. Our data for this binary system are too sparse to significantly constrain its  orbital properties, though our exploration of binary models with \texttt{thejoker} \citep{2017ApJ...837...20P} weakly suggest that the period may be $\sim$15 days or $\sim$30 days. We advocate for continued monitoring of this star to establish its binarity at even higher significance and constrain its period and center-of-mass velocity.

\begin{deluxetable*}{lccccc|cc}
\tabletypesize{\footnotesize}
\tablecaption{Measured Properties for the Complete Sample of 16 \uno{} members}
\tablehead{
\colhead{Object} & \colhead{Gaia DR3 source\_id}  & \colhead{$r_0$} & \colhead{$v_{\mathrm{Epoch\,1}}$ (km s$^{-1}$)} & \colhead{$v_{\mathrm{Epoch\,2}}$ (km s$^{-1}$)} & \colhead{$\lvert\Delta v\rvert / \sqrt{\epsilon_{v_{E1}}^2 + \epsilon_{v_{E2}}^2}$} & \colhead{$(B-V)_0$} & \colhead{LRIS [Fe/H]}}
\startdata
\texttt{S24\_M1} & 4024083571202406912 & 17.72 & 89.2 $\pm$ 1.2 & 89.9 $\pm$ 1.1 & 0.44 & 0.60 $\pm$ 0.03 & $-2.47 \ \pm$ 0.33 \\
\texttt{S24\_M2 \ (Binary**)} & 4024177442007708672 & 18.70 & 81.3 $\pm$ 1.4 & 83.3 $\pm$ 1.2 & 1.07** & 0.40 $\pm$ 0.03 & $-2.87 \ \pm$ 0.33 \\
\texttt{S24\_M3} & 4024177648166139904 & 18.74 & 89.8 $\pm$ 1.4 & 88.8 $\pm$ 1.2 & 0.54 & 0.37 $\pm$ 0.03 & $-2.90 \ \pm$ 0.34 \\
\texttt{S24\_M4 \ (Binary?)} & 4024178472800038016 & 18.90 & 92.5 $\pm$ 1.5 & 88.2 $\pm$ 1.3 & \textbf{2.19} & 0.39 $\pm$ 0.03 & $-2.99 \ \pm$ 0.34 \\
\texttt{S24\_M6} & 4024177648166141184 & 20.11 & 88.8 $\pm$ 2.1 & 90.1 $\pm$ 1.7 & 0.49 & 0.52 $\pm$ 0.04 & $-2.28 \ \pm$ 0.33 \\
\texttt{S24\_M7} & 4024178472800038144 & 20.33 & 86.6 $\pm$ 2.6 & 89.4 $\pm$ 1.8 & 0.87 & 0.52 $\pm$ 0.05 & $-2.75 \ \pm$ 0.35 \\
\texttt{S24\_M9} & 4024177442007709440 & 20.38 & 93.6 $\pm$ 2.7 & 89.4 $\pm$ 1.8 & 1.29 & 0.50 $\pm$ 0.05 & $-2.75 \ \pm$ 0.32 \\
\texttt{S24\_M10} & 4024177751245359744 & 20.39 & 84.3 $\pm$ 2.7 & 86.5 $\pm$ 1.9 & 0.66 & 0.57 $\pm$ 0.05 & $-2.85 \ \pm$ 0.35 \\
\texttt{S24\_M11 (Binary?)} & 4024177682525881344 & 20.44 & 93.5 $\pm$ 2.7 & 86.8 $\pm$ 1.9 & \textbf{2.04} & 0.46 $\pm$ 0.04 & $-2.27 \ \pm$ 0.34 \\
\texttt{S24\_M12} & 4024177442007706752 & 20.43 & 86.2 $\pm$ 3.0 &  &  & 0.49 $\pm$ 0.05 & $-2.86 \ \pm$ 0.33 \\
\texttt{S24\_M13} & 4024177648166141312 & 20.71 & 88.5 $\pm$ 3.5 & 89.1 $\pm$ 2.1 & 0.16 & 0.53 $\pm$ 0.09 & $-2.35 \ \pm$ 0.48 \\
\hline 
\texttt{C25\_M1} & 4024177682525880704 & 20.70 &  & 86.4 $\pm$ 2.0 &  & 0.60 $\pm$ 0.04 & $-2.32 \ \pm$ 0.33 \\
\texttt{C25\_M2 \ (Binary?)} &  & 22.19 &  & 95.5 $\pm$ 2.4 &  &  &  \\
\texttt{C25\_M3} &  & 22.51 &  & 89.2 $\pm$ 4.7 &  &  &  \\
\texttt{C25\_M4} &  & 22.62 &  & 94.4 $\pm$ 10.4 &  &  &  \\
\texttt{C25\_M5} &  & 23.02 &  & 89.7 $\pm$ 7.1 &  &  &  \\
\enddata\tablecomments{Previously-reported members from \citetalias{2024ApJ...961...92S} are shown above the horizontal divider, while new members are shown below that line. Note that the  \citetalias{2024ApJ...961...92S} velocities reported here are updated based on a re-reduction of the original spectra. \texttt{S24\_M2}, highlighted above with a **, is a confirmed binary based on our five epochs of observation (see \tabref{binary}). The other three binaries should be regarded as candidates until more measurements become available. The quoted $r_0$ magnitudes are from Legacy Surveys DR10 \citep{2019AJ....157..168D}.  Note that the quoted uncertainties for [Fe/H] do not include the global zeropoint uncertainty of 0.3 dex that we assumed for \unoshort's mean metallicity.  \label{tab:members}}\end{deluxetable*}
\begin{deluxetable}{l c c}
\tablecaption{Radial Velocity Measurements for the Now-Confirmed Binary Star \texttt{S24\_M2}}
\tablehead{
  \colhead{Instrument} & 
  \colhead{MJD} & 
  \colhead{Velocity (km/s)}  
}
\startdata
DEIMOS 2023 & 60058.29 & 81.3 $\pm \; 1.4$ \\
DEIMOS 2025 & 60767.32 & 83.3  $\pm \;1.2$  \\
MagE & 60799.04 & 90.8 $\pm \;3.7$ \\
GMOS & 60820.36 & 75.5 $\pm \; 4.2$\\
HIRES & 60823.26  &  81.0 $\pm \; 1.8$\ \\
\enddata
\tablecomments{The listed velocity uncertainties include systematic terms. A $\chi^2$ test leads us to reject the null hypothesis of no variability at the $p < 0.05$  level ($p = 0.023$). \label{tab:binary}}

\end{deluxetable}

\subsection{Three New Binary Candidates}
\label{sec:bincands}
To search for additional binary candidates based exclusively on our DEIMOS measurements, we computed the statistical significance of the difference between velocity measurements across epochs, $\lvert\Delta v\rvert/\sqrt{\epsilon_{v_{E1}}^2 + \epsilon_{v_{E2}}^2}$, where $\epsilon_{v}$ is the total velocity error of each star (including the systematic term). The results of these calculations are reported in \tabref{members}.  Within our sample of 10 stars with two epochs of DEIMOS measurements,  exactly two (\texttt{S24\_M4} and \texttt{S24\_M11}) were found to exhibit greater than two-sigma variation, i.e., $\lvert\Delta v\rvert/\sqrt{\epsilon_{v_{E1}}^2 + \epsilon_{v_{E2}}^2} >2$, while the remainder display no strong evidence for variation ($<1.3\sigma$). We therefore flagged both of these stars as candidate binaries. The star \texttt{S24\_M4} was the second member that \citetalias{2024ApJ...961...92S} isolated as having significant influence over their measurement of \unoshort{}'s velocity dispersion; the difference across our DEIMOS measurements thus appears to favor their hypothesis that this star is a binary.
\par Beyond these two candidates, we identified a third member star that is plausibly a binary but for which only a single epoch is available: the new member \texttt{C25\_M2}. This velocity of this star ($v_{\rm Epoch 2} = 95.5  \pm 2.4 ~\kms{}$) is clearly above \unoshort{}'s mean velocity of $v_{\rm sys.} = 89.0~\kms$, and if we generously assumed that \unoshort{}'s true velocity dispersion is $\sigma_v = \vdispFreqCompleteEtwo$~\kms{} (our nominal \textit{upper limit} from the analysis that follows), it would be a $\sim2\sigma$ outlier. Accordingly, we flagged this star as a candidate binary as well, though more epochs are needed to test this hypothesis further. \newline

\par The identification of four binaries (one confirmed + three candidate) in a sample of 16 \unoshort{} stars could imply a binary fraction of $f_b \geq  0.25$. This would be consistent with the (high) binary fraction observed in some low-mass globular clusters \citep{2012A&A...540A..16M,2016MNRAS.455.3009M} and in the small number of ultra-faint dwarf galaxies for which estimates are available \citep{2013ApJ...771...29G,2019MNRAS.487.2961M,2024ApJ...967..165F}. Thus, while this rough limit on \unoshort{}'s  binary fraction does not inform us about the system's classification,  it would not be unreasonable to have four or more binaries in our spectroscopic sample.

\subsection{Velocity Dispersion Constraints}
\label{sec:vdisp}
We determined the velocity dispersion of \uno{} through both a frequentist analysis and a more typical Bayesian fit. In both cases, we adopted the likelihood from \citet{2006AJ....131.2114W}, 
\begin{align}
 \ln (\mathcal{L}) \propto -\frac{1}{2} \sum_{i=1}^N \ln \left(\epsilon_{v,i}^2+\sigma_{v}^2\right)-\frac{1}{2} \sum_{i=1}^N \frac{\left(v_{\rm hel, i}- v_{\rm sys.}\right)^2}{\left(\epsilon_{v,i}^2+\sigma_v^2\right)} 
 \label{eq1}
\end{align}
which includes two free parameters: the systemic mean velocity ($v_{\rm sys.}$) and the intrinsic global velocity dispersion ($\sigma_v$).\footnote{We also explored kinematic models including a linear velocity gradient, finding that our data were insufficient to place meaningful constraints on a possible gradient.} Both analyses were carried out with two subsets of members: a ``Complete Epoch 2'' sample of $N=14$ stars with Epoch 2 measurements in which we excluded the confirmed binary only, and a ``Pure Epoch 2'' sample of $N=11$ stars for which we further removed the three candidate binaries discussed in \secref{bincands}. We did not combine velocity measurements across our two DEIMOS epochs.
\par
\begin{figure*}[!ht]
    \centering
     \includegraphics[width=0.9\textwidth]{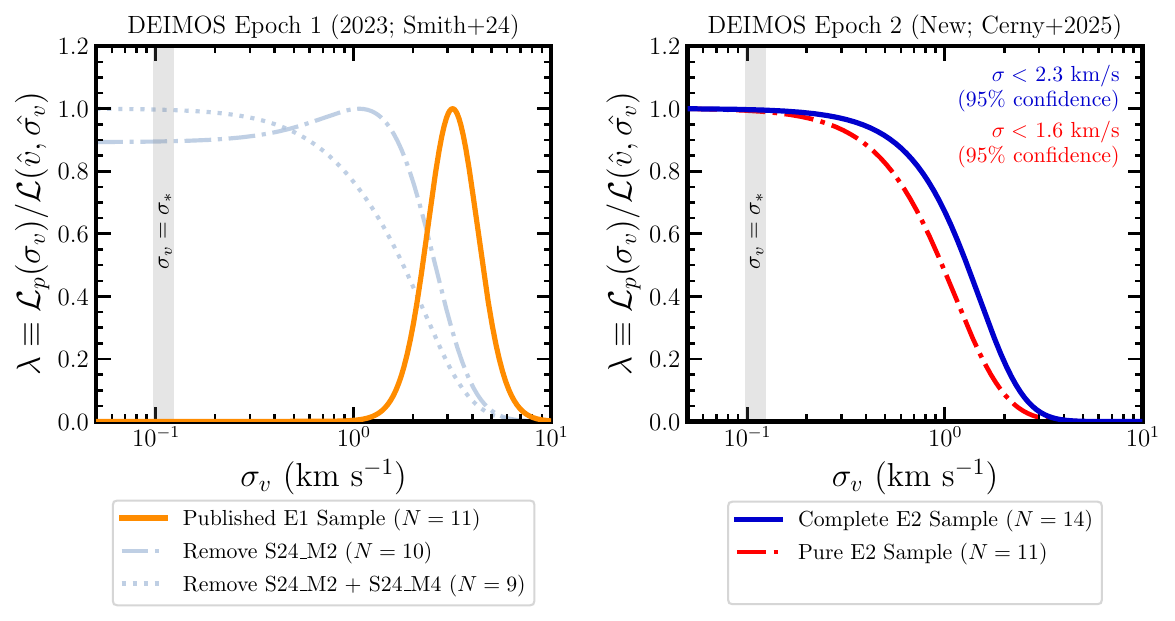}
    \caption{\textbf{Stringent limits on the velocity dispersion of \uno{}}. We display the normalized profile likelihoods for the velocity dispersion ($\sigma_v$) for the re-reduced 2023 DEIMOS observations (left) and the new 2025 DEIMOS observations (right), using different member subsets in each case. The left panel reproduces the conclusions of \citetalias{2024ApJ...961...92S}, demonstrating the clearly-resolved dispersion when all 11 original members were included (solid orange). In the righthand panel, we display our primary constraints from this work for the Complete Epoch 2 (blue) and Pure Epoch 2 (red) samples. The likelihood in both cases is maximized for the smallest allowed dispersions, and the original dispersion of $\sigma_v = 3.7$~\kms{} is clearly ruled out.}
    \label{fig:dynamics}
\end{figure*}

 Our frequentist analysis involved computing the profile likelihood $\mathcal{L}_p(\sigma_v) \equiv \underset{v_{\rm sys.}}{\max} \; \mathcal{L}(\sigma_v, v_{\rm sys.})$ over a logarithmic grid of velocity dispersions defined over  $0.05~\kms \leq~\sigma_v \leq~10~\kms$ for both of our Epoch~2 member sub-samples.\footnote{The choice of $0.05~\kms$ as the lower bound of our grid was motivated by the smallest reported estimate of the dispersion expected in the stellar-only case \citep{2024ApJ...965...20E}, though $\sigma_* \approx 0.1~$\kms{} is likely more realistic \citep{2025MNRAS.539.2485D}.} In other words, we optimized to find the maximum-likelihood estimate of $v_{\rm sys.}$ for each value of $\sigma_v$ -- analogous to integrating over $v_{\rm sys.}$ as a nuisance parameter in the Bayesian case (see e.g., \citealt{sprott2008statistical}). We then further evaluated the profile likelihood ratio $
\lambda(\sigma_v) \equiv \mathcal{L}_p(\sigma_v)/\mathcal{L}(\hat{v},\hat{\sigma_v}) $
across our $\sigma_v$ grid, where the denominator is the global maximum likelihood estimate from simultaneous optimization of both parameters.  The resulting profile likelihood ratio curves are shown in the righthand panel of \figref{dynamics} for both Epoch 2 member subsamples; for ease of comparison, we also present the analogous curves derived from the original 2023 data in the lefthand panel.
\par For the Complete Epoch 2 sample (solid blue line in \figref{dynamics}), we found that the likelihood is maximized for $\leq 0.1$~\kms{} and relatively flat out to $0.6~\kms$ before declining to become statistically negligible for $\sigma_v \gtrsim  4~\kms$. The Pure Epoch 2 sample (red dash-dotted line) similarly begins as flat for the smallest dispersions,  but more quickly declines to become negligible at $\sigma_v \gtrsim  3~\kms$. The flatness of the profile likelihoods for the smallest dispersions reflects the limited constraining power of our dataset to dispersions significantly below the per-star velocity uncertainty floor of $1.1~\kms$. For the Complete Epoch 2 sample, there is a 121:1 likelihood ratio between the maximum-likelihood estimate of $\sigma_v = \sigma_* \approx 0.05$--$0.1~$\kms{} and the $3.7$~\kms{} dispersion reported by \citetalias{2024ApJ...961...92S}, providing strong evidence against the original high measurement and positive evidence for a small or absent dark matter content. The evidence against a large velocity dispersion grows even stronger if we instead consider the Pure Epoch 2 sample; in that case, we find a 281:1 likelihood ratio when making the same comparison. 

\par To formalize our frequentist constraints, we derived 95\% confidence upper limits on the velocity dispersion based on the likelihood ratio test approach described in Section 2.4 of \citet{2011EPJC...71.1554C}. These limits were calibrated through toy Monte Carlo simulations to ensure proper coverage; see Appendix \ref{sec:appBfrequentist} for a complete description of our procedure. This yielded a tight 95\% confidence level  upper limit of $\sigma_v < \vdispFreqCompleteEtwo$~\kms{} for the Complete Epoch 2 sample and an even tighter limit of $\sigma_v < \vdispFreqPureEtwo$~\kms{} for the Pure Epoch 2 sample. These limits can again be compared to the original measurement of $\sigma_v = 3.7^{+1.4}_{-1.0}$~\kms{} from \citetalias{2024ApJ...961...92S}. This substantial change in the dispersion reflects a combination of the removal of the confirmed binary, our larger sample size, and the regression to the mean of member velocities when observed at higher precision. We conservatively adopt the first of these limits as our primary measurement for this work, which avoids unduly rejecting stars with velocity measurements that could reflect random $2\sigma$ fluctuations given the observational uncertainties.

\par For our complementary Bayesian approach, we derived posterior probability distributions for \unoshort{}'s mean velocity and velocity dispersion through dynamic nested sampling with \texttt{dynesty} \citep{2020MNRAS.493.3132S}. Assuming uniform priors of $\rm -500~\kms{} < v_{\rm sys.} < 500~\kms{}$ and $0~\kms{} \rm  < \sigma_v  < 10~\kms{}$. From the marginalized (1D) posteriors derived based on our fit to the Complete Epoch 2 sample, we determined an updated systemic mean velocity of $v_{\rm sys.} = 89.0\,\pm\,0.6$~\kms{} and a velocity dispersion upper limit of $\sigma_v < \vdispBayesCompleteEtwo$~\kms{} at the 95\% credible level.  If we instead adopted a log-uniform prior of $-1 < \rm \log_{10}(\sigma_v) < 1$, which is physically reasonable given the lower bound of $\sigma_* \approx 0.1$~\kms{} expected from the stellar-only scenario, this upper limit tightened to $\sigma_v < \vdispBayesLogCompleteEtwo$~\kms{} at the 95\% credible level. These estimates bracket the frequentist result, and their divergence reflects the significant impact of the form of the assumed Bayesian prior.
Carrying out the same procedures for the Pure Epoch 2 sample, we found upper limits of $\sigma_v < \vdispBayesPureEtwo$~\kms{} (uniform prior) and $\sigma_v < \vdispBayesLogPureEtwo$~\kms{} (log-uniform prior). This latter result is the strongest possible constraint that can be derived based on standard priors applied to the Epoch 2 data. 
\par Importantly, we found that the two-parameter models assumed for these fits were disfavored relative to a no-dispersion model. Comparing the Bayesian evidence provided by our \texttt{dynesty} fits to the evidence computed from similar fits where $\sigma_v$ was set to zero (i.e., no spread), we determined Bayes factors of  $2\ln\beta = \BayesFactorCompleteEtwo$ and $2\ln\beta = \BayesFactorPureEtwo$ for the Complete and Pure Epoch 2 samples, respectively (assuming the original uniform prior on $v_{\rm sys.}$). These values indicate a ``positive'' preference for the null hypothesis that there is no intrinsic velocity spread in the system \citep{Kass01061995}.  
\par In summary, we report a nominal 95\% confidence level upper limit of $\sigma_v < \vdispFreqCompleteEtwo$~\kms{} for \unoshort{}'s velocity dispersion from our frequentist analysis, with a complete range of upper limits spanning $1.4$~\kms{} to $\vdispBayesCompleteEtwo$~\kms{} depending on the member subset and the statistical approach applied. Both likelihood-ratio-based and Bayes-factor-based model comparisons favor small or negligible velocity dispersions.
\subsection{Dynamical Mass, Mass-to-Light Ratio, and Dark Matter Content}
\par Using the new velocity dispersion measurements from above, we estimated \unoshort{}'s dynamical mass with the mass estimator from \citet{2010MNRAS.406.1220W}. Taking \unoshort{}'s azimuthally-averaged half-light radius of $r_{1/2} = 2.1$~pc\footnote{The azimuthally-averaged half-light radius is defined as $r_{1/2} \equiv a_{1/2}\sqrt{1-\epsilon}$, where $\epsilon \equiv 1 - \frac{b}{a}$ is the ellipticity. For \citetalias{2024ApJ...961...92S}'s estimate of $a_{1/2} = 3$~pc and $\epsilon = 0.5$, this yields $r_{1/2} = 2.1$~pc.} and our velocity dispersion constraint of $\sigma_v < \vdispFreqCompleteEtwo$~\kms{}, we found a dynamical mass within the half-light radius of $M_{1/2} \lesssim \WolfMassFreqCompleteEtwo \; M_{\odot}.$ Assuming an enclosed luminosity of $L_V(r < r_{1/2}) = 5.5 \; L_{\odot}$ \citepalias{2024ApJ...961...92S}, the corresponding mass-to-light ratio for \unoshort{} is $M/L_V \lesssim \MLRatioFreqCompleteEtwo \; \rm M_{\odot}L_{\odot}^{-1}.$ If instead we used the tighter dispersion derived limit from the frequentist analysis of the Pure Epoch 2 sample ($\sigma_v < \vdispFreqPureEtwo$~\kms), we obtained $M_{1/2} \lesssim \WolfMassFreqPureEtwo \; M_{\odot}$ and $M/L_V \lesssim \MLRatioFreqPureEtwo \; \rm M_{\odot}L_{\odot}^{-1}$. Finally, if we used the strongest dispersion limit quoted above,  $\sigma_v < \vdispBayesLogPureEtwo$~\kms{}  (from the log-prior Bayesian analysis of the Pure Epoch 2 sample), these limits drop to $M_{1/2} < \WolfMassBayesLogPureEtwo \; M_{\odot}$ and $M/L_V \lesssim \MLRatioBayesLogPureEtwo \; \rm M_{\odot}L_{\odot}^{-1}$.\footnote{Note that all of these mass and mass-to-light ratio estimates intentionally exclude the (substantial) uncertainty contributions from $L_V$ and $r_{1/2}$.}   Based on these new estimates, we unequivocally conclude that the dark matter content of \unoshort{} is smaller than initially reported. This is true even when considering the results derived from the member subsamples proposed by \citetalias{2024ApJ...961...92S}; for example, they found that removing the confirmed binary yielded  $M/L_V = 1900^{+4400}_{-1600} \rm \ M_\odot L_\odot^{-1}$. 
\par The key outstanding question is whether there is \textit{any} dark matter in the system: while the new kinematic data favor velocity dispersions suggestive of the absence of dark matter, our formal mass-to-light ratio limits permit the possibility of substantial unseen mass in the system. We therefore stress that we cannot -- and do not -- claim there is no dark matter in \unoshort{}. Instead, the data support the narrower conclusion that \textit{there is no positive observational evidence for the presence of dark matter in \unoshort{}}. This latter statement motivates our critical re-examination of the system's classification in \secref{classification}, as the stellar kinematics of \unoshort{} -- previously taken as the motivating evidence for a galaxy classification for the system -- no longer suggest that the system is a galaxy.
\par One further implication of these new dynamical measurements is that the system is now a less appealing target for the indirect detection of dark matter annihilation through its $\gamma$-ray signature. Based on the scaling relation from  \citet{2019MNRAS.482.3480P}, we report an approximate 95\% confidence level $J$-factor limit of $\log_{10}(J(0.5^\circ)) < \JfactorFreq$. This is an an order of magnitude smaller than prior estimates based on the \citetalias{2024ApJ...961...92S} data, which have converged around $\log_{10}(J) \approx  21$ for velocity-independent ($s$-wave) annihilation \citep[e.g.,][]{2024ApJ...965...20E,2024ChPhC..48k5112Z}. While we strongly advise that the system be excluded from indirect detection analyses because of our ultimate conclusion that \unoshort{} is more likely a star cluster,  we note that our revised limit remains larger than any known ultra-faint dwarf galaxy $J$-factor. This underscores the immense value of pursuing stellar kinematics measurements for nearby, compact satellites given the  scaling of the annihilation signal as $\log_{10}(J)\propto D_\odot^{-2}$ (see e.g., \citealt{2025ApJ...978L..43C}).

\begin{figure*}[!ht]
    \centering
     \includegraphics[width=\textwidth]{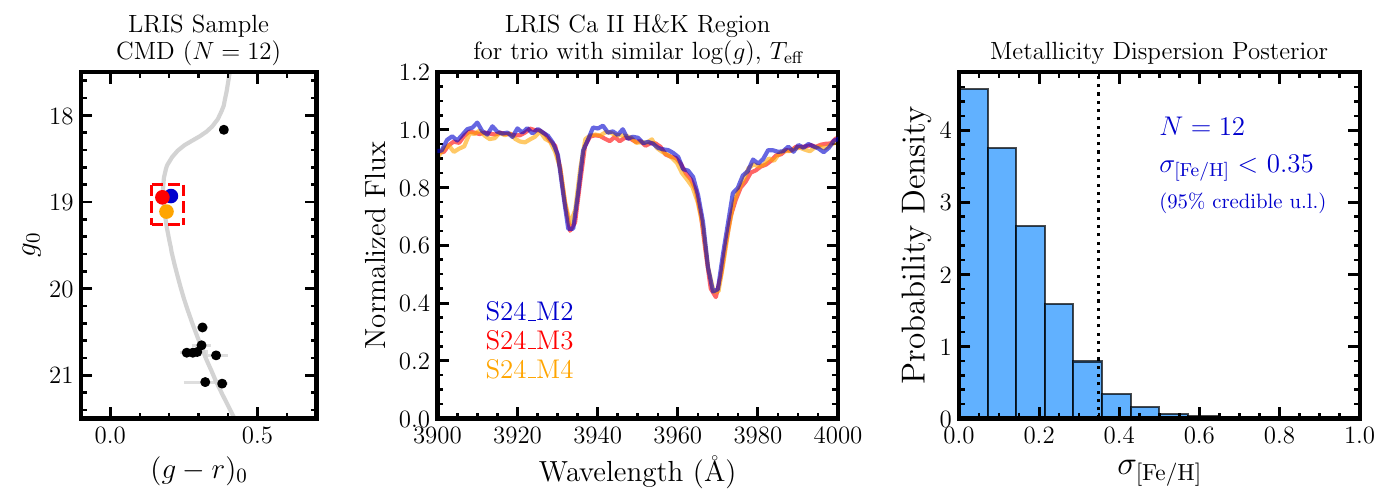}
    \caption{\textbf{The absence of a significant metallicity spread in  \unoshort{}.} (Left) Color--magnitude diagram of the 12 \unoshort{} stars observed with Keck/LRIS (using PanSTARRS-1 photometry). We box three stars with similar stellar parameters that are useful for tests of chemical homogeneity. (Center) The LRIS spectra for the three boxed stars (\texttt{S24\_M2},\texttt{S24\_M3}, and \texttt{S24\_M4}) covering the Calcium H and K lines. The three spectra are very similar, suggesting similar metallicities for these three stars given their similar surface gravities and temperatures ([Fe/H] = $-2.87 \pm 0.33$, $-2.90 \pm 0.34$, $-2.99 \pm 0.34$). See \figref{all_lris_spectra} for the LRIS spectra of the remaining nine stars. (Right) Marginalized posterior probability distribution on the intrinsic metallicity dispersion of \unoshort{} $(\sigma_{\rm [Fe/H]}$) derived from the complete LRIS sample of 12 member stars. } 
    \label{fig:chemistry}
\end{figure*}

\section{No Evidence for a Large Stellar Metallicity Spread in UMaIII/U1}
\label{sec:chemistries}
\subsection{Keck/LRIS Observations and Calcium K Metallicities}
\label{sec:fehdata}
\par To chemically test the true nature of \unoshort{} through its mean metallicity and metallicity dispersion, we obtained low-resolution ($R \approx 1000$), optical ($\sim 3750$--$9000 \; \rm \AA$) spectra of its 12 brightest member stars with the Low-Resolution Imaging Spectrometer (LRIS; \citealt{1995PASP..107..375O}) on Keck I on UTC 2025-02-04 and 2025-02-05. The immediate goal of these observations was to acquire high signal-to-noise spectra covering the Calcium II K absorption feature at $\sim 3930\rm \AA$ from which we could derive equivalent-width-based metallicities. We thus exclusively made use of the data from the instrument's blue arm \citep{1998SPIE.3355...81M} for which we used  the 600/4000 grism.  \par We observed \unoshort{} for a total of $5.8$~hours through a single multi-slit LRIS mask (with 1.0$\arcsec$ slits) covering the 11 original \citetalias{2024ApJ...961...92S} members in addition to a twelfth star, \texttt{C25\_M1}, that we later confirmed as a velocity member with DEIMOS. This represents a complete sample of all suspected members of the system down to $g_0\sim 22$ (see the color--magnitude diagram in the lefthand panel of \figref{chemistry}). Exposure times were typically 30 minutes, with calibrations taken after every two blue-arm science exposures to account for instrument flexure. The LRIS data were subsequently
reduced with the standard utilities in \texttt{PypeIt}. We show representative LRIS spectra for three of our members in \figref{chemistry} and display the full sample in Appendix \ref{sec:appCLRISspectra}, \figref{all_lris_spectra}.
\par We determined iron abundances ([Fe/H]) from our LRIS data based on the KP index calibration from \citet{1999AJ....117..981B}, which is ideally suited for our low-resolution spectra given its decades-long heritage with even-lower-resolution prism surveys.  This calibration uses pseudo-Equivalent Widths (EWs) of the Ca II K line paired with $B-V$ colors as a temperature proxy. Here, we followed a similar procedure to \citet{2018ApJ...856..142C} and \citet{2025arXiv250616462L}. As described there, each spectrum was first velocity corrected and continuum normalized. Pseudo-EWs were then derived within the three wavelength windows specified by \citet{1999AJ....117..981B}. To capture continuum placement uncertainty in the EW error budget, we then Monte Carlo-resampled each observed spectrum 5000 times, re-normalized, and then re-computed the KP index, yielding a distribution of pseudo-EW measurements from which we derived a single estimate and uncertainty for each star.
To obtain $B-V$ colors, we queried PanSTARRS~1 DR2 photometry \citep{2016arXiv161205560C} using the Astro Data Lab \citep{2014SPIE.9149E..1TF,2020A&C....3300411N}, which we transformed into  $B$ and $V$ magnitudes using the relations from \cite{2012ApJ...750...99T}. Comparing to carefully-calibrated Stetson $B,V$ photometry of the metal-poor globular clusters M15 and M92 \citep{2019MNRAS.485.3042S}, we found that the PS1-based $B-V$ colors were overestimated by 0.03 mag with a sigma-clipped scatter of 0.03~mag; we therefore applied a -0.03 mag correction and added a 0.03 mag uncertainty floor in quadrature to our color estimates. Finally, we subtracted a reddening of $E(B-V) = 0.02$ for each star based on the maps of \citet{2011ApJ...737..103S}.

\par  The total [Fe/H] uncertainty for each star was calculated by combining three sources of error in quadrature: the spectroscopic uncertainty in the Ca II K EW measurements, uncertainty in the $(B-V)_0$ colors, and uncertainties on the \citet{1999AJ....117..981B} KP calibration. Because of the very high signal-to-noise of our LRIS data ($S/N > 25$ for all stars), the spectroscopic term is small for each star ($0.05$--$0.15$~dex) -- even after our Monte Carlo procedure that accounts for the continuum placement uncertainty. The dominant source of uncertainty for our sample is therefore the KP calibration uncertainty, which contributes $\sim$0.25--0.3 dex to the error budget for each star. This relatively large contribution arises from the small number of calibrator stars at low KP index and low (bluer) B-V colors in the empirical relation from \citet{1999AJ....117..981B}. We take these substantial calibration uncertainties at face value, though reducing this contribution is an area of ongoing work.

\par Finally, we note that \citet{2025arXiv250616462L} found that the KP calibration from  \citet{1999AJ....117..981B} yielded metallicities 0.29 dex too low relative to more modern metallicity scales based on a comparison to both medium-resolution Calcium Triplet spectra and high-resolution iron-line measurements. Our choice above to re-zero our $B-V$ colors naturally shifts our derived LRIS metallicities in the same direction by at least $0.1$~dex, motivating us to forego this 0.29 dex global offset. To capture any remaining differences in metallicity scale, we later adopted a 0.3 dex global zeropoint error on our mean metallicity measurement (see subsection below). We stress that revisions to this global zeropoint would shift all stars equally and does not affect the \textit{relative} metallicities between stars, which are more important for understanding the system's classification.
\par Our complete set of LRIS [Fe/H] measurements are provided in \tabref{members}. These metallicities span a relatively narrow range of $\rm [Fe/H] = -2.27 \pm 0.34$ to $\rm [Fe/H] = -2.99 \pm 0.34$. Notably, the maximum metallicity difference between any two stars is just $\sim 1.5\sigma$.

\subsection{Mean Metallicity and Metallicity Dispersion Constraints}
We determined \unoshort{}'s mean metallicity ([Fe/H]) and intrinsic metallicity dispersion ($\sigma_{\rm [Fe/H]}$) through a Bayesian fit mimicking our analysis for the velocity dispersion above. We assumed a Gaussian likelihood of the same form as Equation~\ref{eq1} and fit the complete LRIS metallicity dataset comprised of $N=12$ stars; this includes the confirmed binary member. The only salient differences relative to our Bayesian velocity dispersion analysis were the priors for the \texttt{dynesty} nested sampling. Specifically, we instead adopted a wide uniform prior of $-4 < \rm [Fe/H] < 0$  for \unoshort{}'s mean metallicity and a uniform prior of $0\rm \ dex< \sigma_{\rm [Fe/H]} < 1 \ dex$ for its metallicity dispersion.
\par These fits demonstrated that  \unoshort{} is a low-metallicity stellar system with no evidence for large internal metallicity variations. 
From the marginalized 1D posterior for each parameter, we found a mean metallicity of $\rm [Fe/H]$ = \BayesianFeH{} (stat.) with an upper limit of $\sigma_{\rm [Fe/H]} < \BayesianFeHDispUL$~dex at the 95\% credible level; see \figref{chemistry} for the latter parameter's posterior. To the statistical  uncertainty on the mean metallicity, we further added a 0.3 dex global zeropoint uncertainty to account for possible offsets against recent Calcium-Triplet-based measurements, yielding a final measurement of $\rm [Fe/H]$~=~\BayesianFeH{} (stat.) $\pm~ 0.3$~(zeropoint). The Bayes factor relative to models with $\sigma_{\rm [Fe/H]}$ fixed to zero was found to be $2\ln\beta = \BayesFactorFeH$, indicating a mildly significant positive preference for the null hypothesis of no intrinsic metallicity dispersion in the system. If we instead performed the same analysis with a physically-motivated log-uniform prior of $-2 < \log_{10}(\sigma_{\rm [Fe/H]}) < 0$, the mean metallicity was unchanged but the dispersion upper limit dropped to $\sigma_{\rm [Fe/H]} < 0.22$  at the 95\% credible level;  the corresponding Bayes factor was $2\ln\beta = -1.0$.
\par We explored the possibility that our non-detection of a metallicity spread could arise from overestimated [Fe/H] uncertainties, but ultimately found that our primary 95\% credible level upper limit was consistent at the $\sim$0.02--0.03~dex level even if our uncertainties were overestimated by as much as 40\%; in that limiting case, the Bayesian evidence would still favor the no-spread model ($2\ln\beta < 0$). This is a consequence of the narrow observed range in stellar metallicities in the system, which rules out models with $\sigma_{\rm [Fe/H]} \gg 0.35$~dex irrespective of the per-star measurement uncertainties. A more stringent constraint on \unoshort{}'s metallicity dispersion would require reducing the KP calibration uncertainty, which is the subject of ongoing work.

\par Thus, in short, we find that \textit{there is no evidence for substantial internal metallicity variations in \uno{}.} This property suggests that \unoshort{} is more similar to nearly-chemically-homogeneous globular clusters ($\sigma_{\rm [Fe/H]} \lesssim 0.05$--$0.1$~dex; \citealt{2019ApJS..245....5B}) than to self-enriched ultra-faint dwarf galaxies with more extended star formation histories (typically $\sigma_{\rm [Fe/H]} =  0.3$--$0.7$~dex; \citealt{2023ApJ...958..167F,Pace2024arXiv241107424P}; see also \citealt{2012AJ....144..183L,2012AJ....144...76W} and the following section), though the latter scenario is not ruled out.

\begin{figure*}[!ht]
    \centering
    \includegraphics[width=0.95\textwidth]{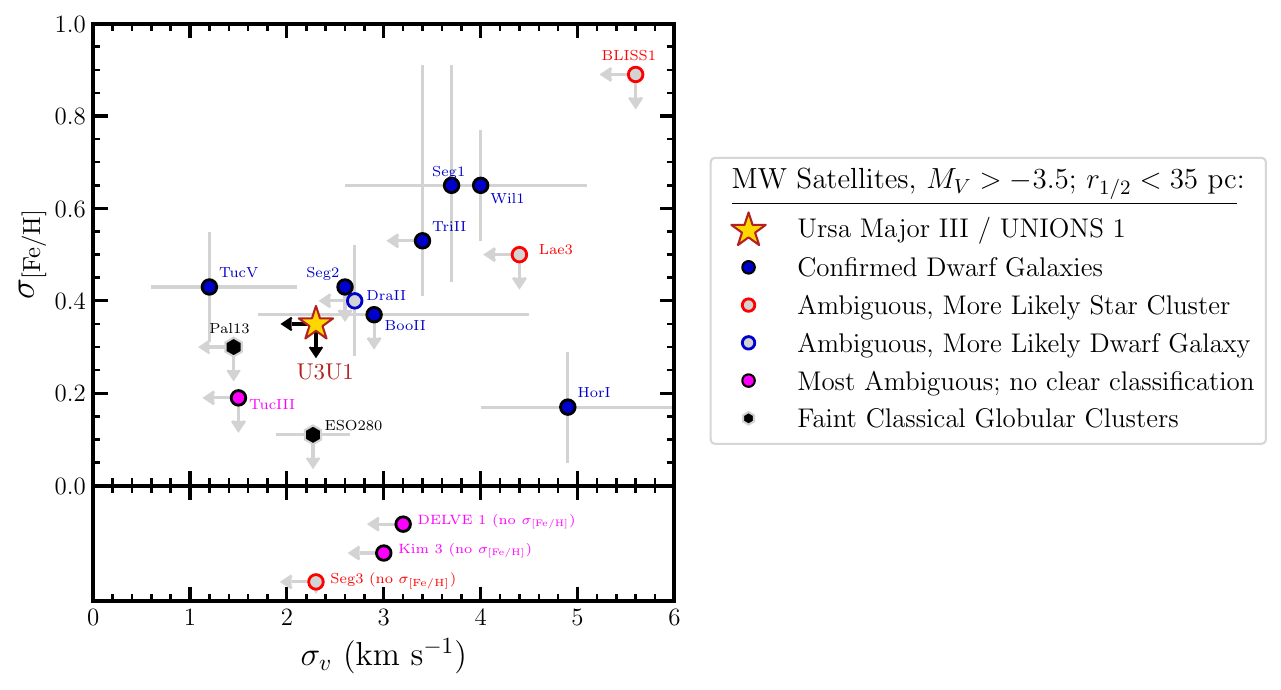}
    \caption{\textbf{\unoshort{}'s low velocity dispersion limit and metallicity dispersion limit in the context of the faint, compact Milky Way satellite population.} We compare our new measurements for the system to a carefully curated sample of systems with $M_V > -3.5$; $r_{1/2} < 25$~pc for which there are both $\sigma_{\rm [Fe/H]}$ and $\sigma_v$ measurements available (top sub-panel) or only a  $\sigma_v$ measurement (bottom sub-panel); we exclude systems with possibly contaminated/impure stellar memberships. These include confirmed dwarf galaxies (filled blue), ambiguous compact systems (unfilled red or blue, for systems more likely to be clusters or dwarfs, respectively; magenta if entirely unclear), and select globular clusters (black). Our new limits for \unoshort{} rank among the strongest of any ultra-faint system in this size and luminosity regime; the only system with stronger limits for both types of dispersion is Tucana~III. See Appendix \ref{sec:appDcomparison} for a discussion of the comparison sample (and associated references). }
    \label{fig:summary}
\end{figure*}

\section{Reclassifying UM\lowercase{a}III/U1 as a Likely Star Cluster}
\label{sec:classification}
\citet{2012AJ....144...76W} defined a galaxy as a ``gravitationally bound collection of stars whose properties cannot be explained by a combination of baryons and
Newton’s laws of gravity,'' which, within the concordance cosmological framework, can be interpreted as stating that galaxies are distinguished from star clusters by the presence of dark matter. They identified two primary empirical diagnostics that together positively identify galaxies: large mass-to-light ratios, which directly quantify the presence of unseen matter not accounted for by the luminous stellar component, and significant metallicity dispersions, which point to multiple generations of star formation and the presence of a potential well deep enough to retain supernova ejecta. 
\par \citetalias{2024ApJ...961...92S}'s classification of \unoshort{} as a galaxy candidate rested entirely on the first diagnostic -- the system's large velocity dispersion and high implied mass-to-light ratio. 
The new data presented here removes this pivotal piece of evidence, supplanting it with a strong velocity dispersion limit ($\sigma_v < \vdispFreqCompleteEtwo$~\kms{} and potentially $\sigma_v < \vdispFreqPureEtwo$~\kms , at the 95\% confidence level) and a statistical preference for a negligible velocity spread among \unoshort{}'s stellar members. As for the second diagnostic, we have shown that the system lacks a large metallicity dispersion ($\sigma_{\rm [Fe/H]}< 0.35$~dex, at the 95\% credible level) and that there is again a statistical preference for zero metallicity spread (from our Bayes factor comparisons). Collectively, these results suggest there is no longer any positive observational evidence in support of a galaxy classification for \unoshort{}. We therefore re-classify the system as a likely star cluster. Under this interpretation, \unoshort{} is by far the least-luminous ancient star cluster yet discovered, and its long-term stability may require that $>50\%$ of the system's baryonic mass is in unseen white dwarfs, neutron stars, and/or black holes \citep{2025MNRAS.539.2485D,Rostami-Shirazi_2025}. This would also place \unoshort{} among the most metal-poor Milky Way globular clusters known.

\par This reclassification of \unoshort{} as a probable star cluster is broadly supported by a comparison to the existing body of spectroscopic constraints for faint, compact Milky Way satellites  (\figref{summary}). Comparing the system to a carefully curated sample of satellites with $M_V > -3.5$; $r_{1/2} < 35$~pc,  \unoshort{}'s velocity dispersion and metallicity dispersion limits are particularly tight. While a very small number of dwarfs in this regime have a dispersion of \textit{either} type below our upper limits, there is only one claimed ultra-faint dwarf galaxy candidate with stronger limits on \textit{both} $\sigma_v$ and $\sigma_{\rm [Fe/H]}$. This system is Tucana~III,  for which the chemical and dynamical evidence for a galaxy classification is especially uncertain due to the satellite's $r$-process enhancement and significant tidal disruption (respectively). In fact, a recent re-examination of the evidence by \citet{2025ApJ...987..217Z} suggests the system may instead be a star cluster (see also Appendix \ref{sec:appDcomparison}).  The only reliably-confirmed faint, compact dwarf with a smaller velocity dispersion reported is Tucana~V, with a velocity dispersion quoted at $\sigma_v= 1.2^{+0.9}_{-0.6}$~\kms{} \citep{2024ApJ...968...21H};  however, even they note that this dispersion is marginally resolved and opt to also quote an upper limit of $\sigma_v < 3.1$~\kms{} at the 95.5\% level. As  for its abundances, Tucana~V displays remarkable chemical diversity, with a $>1$~dex range in stellar metallicities and dramatic [C/Mg] variation -- very much unlike \unoshort{}. Finally, the confirmed dwarf galaxy Horologium I displays a metallicity dispersion of $\sigma_{\rm [Fe/H]} = 0.17^{+0.20}_{-0.03}$~dex \citep{2015ApJ...811...62K} that nominally falls below our limit for \unoshort{}; however, this measurement is from just five member stars, and the large upper error permits much larger dispersions. Hor I's velocity dispersion of $\sigma_v = 4.8^{+2.8}_{-0.9}$~\kms{} is large and clearly resolved. Thus, there are no clear examples of confirmed faint, compact dwarf galaxies with stronger $\sigma_v$ or  $\sigma_{\rm [Fe/H]}$ limits, supporting the notion that \unoshort{} is distinct from this population and more similar to the faintest globular clusters.

\par Despite all of the above, could \unoshort{} still be a dwarf galaxy? Possibly yes: even our strongest upper limit permits dark matter contents up to $M/L_V < \MLRatioBayesLogPureEtwo \ \rm  M_{\odot}L_{\odot}^{-1}$. Moreover, as described above, there may be ultra-faint dwarfs with smaller claimed metallicity dispersions than our upper limits, and there are no existing theoretical predictions quantifying the stellar metallicity spreads (or lack thereof) expected from a self-enriched galaxy of such low mass. But, in the absence of ultra-precise ($\sigma \lesssim 0.1 \rm \ km \ s^{-1}$) velocity measurements for a large sample of stars across many epochs, it is possible that we may never be able to place the upper limit of $M/L_V \lesssim 2$ necessary to kinematically rule out dark matter beyond any reasonable doubt and thereby exclude the possibility that the system is a galaxy. This is particularly true if the system's intrinsic velocity dispersion is inflated due to primordial binaries (see \citealt{2025MNRAS.539.2485D}). We therefore believe it is most appropriate to regard ultra-faint, compact Milky Way satellites with similarly strong limits as star clusters until clear positive evidence proves otherwise. This avoids the risk of prematurely deriving constraints about dark matter and galaxy formation physics from objects with uncertain classifications.\footnote{To emphasize this point, we draw analogy to the case of faint globular cluster Palomar 13 -- initially thought to contain dark matter before radial velocity monitoring established its velocity dispersion to be the product of initially-unidentified binaries \citep{2002ApJ...574..783C,2004A&A...419..533B,2011ApJ...743..167B}.}

\par To further test the true nature of \unoshort{}, we advocate that future efforts center around seeking evidence of chemical abundance differences (other than iron) across member stars that might re-establish its viability as a galaxy. Our low-resolution LRIS spectra could potentially constrain carbon and magnesium; we currently find these are less constraining than the Calcium K metallicities, but a more detailed study is forthcoming. Neutron-capture abundance measurements for even a single star would also provide valuable information about the system's classification, given that most ultra-faint dwarf galaxy stars are deficient in, e.g., strontium and barium relative to globular cluster stars at similar metallicity \citep{Kirby2017,2019ApJ...870...83J}. Finally, seeking signatures of stellar mass loss such as tidal tails or a depleted stellar mass function could potentially distinguish between the dark-matter-dominated dynamical scenario proposed by \citet{2024ApJ...965...20E} and the collisional, stellar-remnant-dominated scenario discussed by \citet{2025MNRAS.539.2485D}.

\section{Summary}
\label{sec:summary}
We have presented a second epoch of radial velocity measurements for stars in the faintest known Milky Way satellite, \uno{}. Our results remove the previous observational evidence for a large velocity dispersion and dark matter content in the system based on a combination of (1) the confirmation and subsequent exclusion of a stellar binary, (2) regression to the mean of member star velocities when observed at  higher precision, and (3) the identification of five new spectroscopic members. Instead, the data now suggest that \unoshort's velocity dispersion is unresolved and best described by a dispersion close to, or at the level of, the dispersion expected if the system is is entirely comprised of stars. This is particularly true if three newly-identified binary candidates are removed from the member sample.
\par Independently, we have presented the first metallicities for stars in \unoshort{}. From a sample of 12 stars with low-resolution spectra covering the Calcium II K line, we found the system's mean metallicity to be $\rm [Fe/H]$~= \BayesianFeH{}  (stat.)~$\pm~0.3$ (zeropoint). We further found no evidence for large metallicity variations among member stars, with an upper limit on the metallicity dispersion of $<0.35$~dex at the 95\% credible level. The undetectable metallicity and velocity dispersions tilt the balance toward the scenario in which the system is a baryon-dominated, mono-metallicity star cluster. While these limits are not strong enough to conclusively rule out the possibility that the system is a dwarf galaxy, the lack of any single piece of positive observational evidence clearly pointing toward this scenario suggests the star cluster scenario should be treated as the status quo moving forward.

In the context of the broader spectroscopic campaign to understand the smallest and faintest Milky Way satellites, the chemodynamical dataset presented here is the richest available for any system at the extremely low-luminosity frontier ($M_V > 0$) and represents the first demonstration of using low-resolution Calcium II K spectroscopy as a technique for classifying dwarf galaxy candidates. While this dataset has provided clarity into \unoshort{}'s true nature, our work nonetheless highlights the significant challenge in reaching classifications for the least-massive satellites. We were able to assemble this wealth of evidence for \unoshort{} because of its unique status as the closest known ultra-faint Milky Way satellite ($D_{\odot}= 10 \! \pm \! 1$~kpc) and due to a large investment of time targeting a single satellite ($\sim 11.5$ hours of integration on a 10-meter telescope, excluding overheads and calibrations). In the era of the Vera C. Rubin Observatory and \textit{Euclid}, large numbers of similar or slightly more massive systems will likely be discovered far into the outer halo where there is little hope of conducting the type of spectroscopic campaign executed here. We thus urge caution 
as we enter this exciting new era, and forewarn that a complete understanding of the faintest satellites discovered in these surveys will require a spectroscopic campaign of unprecedented scale. Satellites such as \uno{} may therefore be prime candidates for study with upcoming 30m-class telescopes.

\section*{Acknowledgements}
\par We thank the anonymous referee for their constructive feedback on our manuscript.

\par We thank the staffs of the W.M. Keck Observatory, Las Campanas Observatory, and Gemini North for their invaluable assistance in the preparation and execution of our program, especially Michael Lundquist, Percy Gomez, Heather Hershley, and Kristin Chiboucas.

\par We thank Bur\c{c}in Mutlu-Pakdil and the organizers of ``Galactic Frontiers II: Dwarf Galaxies in the Local Volume and Beyond'' at Dartmouth University during which many conversations about \uno{} were held. We thank Alex Drlica-Wagner, Chin Yi Tan, and Scot Devlin for useful discussions thereat.
\par We wish to recognize and acknowledge the very significant cultural role and reverence that the summit of Maunakea has always had within the Native Hawaiian community. We are most fortunate to have the opportunity to conduct observations from this mountain.

\par W.C. gratefully acknowledges support from a Gruber Science Fellowship at Yale University. This material is based upon work supported by the National Science Foundation Graduate
Research Fellowship Program under Grant No. DGE2139841.  Any opinions, findings, and conclusions or recommendations expressed in this material are those of the author(s) and do not necessarily reflect the views
of the National Science Foundation.
D.S.B. and A.P.J. acknowledge support from the National Science Foundation under grant AST-2307599. A.P.J. acknowledges support from the Alfred P. Sloan Foundation.

\par This work was supported by a NASA Keck PI Data Award, administered by the NASA Exoplanet Science Institute (program 2025A-N104). Some of the data presented herein were obtained at Keck Observatory, which is a private 501(c)3 non-profit organization operated as a scientific partnership among the California Institute of Technology, the University of California, and the National Aeronautics and Space Administration. The Observatory was made possible by the generous financial support of the W. M. Keck Foundation. 

\par This research has made use of the Keck Observatory Archive (KOA), which is operated by the W. M. Keck Observatory and the NASA Exoplanet Science Institute (NExScI), under contract with the National Aeronautics and Space Administration.

\par This paper includes data gathered with the 6.5 meter Magellan Telescopes located at Las Campanas Observatory, Chile. 

\par This work is based on observations obtained as part of Gemini program GN-2025A-FT-110 at the international Gemini Observatory, a program of NSF NOIRLab, which is managed by the Association of Universities for Research in Astronomy (AURA) under a cooperative agreement with the U.S. National Science Foundation on behalf of the Gemini Observatory partnership: the U.S. National Science Foundation (United States), National Research Council (Canada), Agencia Nacional de Investigaci\'{o}n y Desarrollo (Chile), Ministerio de Ciencia, Tecnolog\'{i}a e Innovaci\'{o}n (Argentina), Minist\'{e}rio da Ci\^{e}ncia, Tecnologia, Inova\c{c}\~{o}es e Comunica\c{c}\~{o}es (Brazil), and Korea Astronomy and Space Science Institute (Republic of Korea). We have made use of the Gemini Observatory Archive at NSF NOIRLab \citep{2017ASPC..512...53H}.

\par This research uses services or data provided by the Astro Data Lab, which is part of the Community Science and Data Center (CSDC) Program of NSF NOIRLab. NOIRLab is operated by the Association of Universities for Research in Astronomy (AURA), Inc. under a cooperative agreement with the U.S. National Science Foundation.

\par The Pan-STARRS1 Surveys (PS1) and the PS1 public science archive have been made possible through contributions by the Institute for Astronomy, the University of Hawaii, the Pan-STARRS Project Office, the Max-Planck Society and its participating institutes, the Max Planck Institute for Astronomy, Heidelberg and the Max Planck Institute for Extraterrestrial Physics, Garching, The Johns Hopkins University, Durham University, the University of Edinburgh, the Queen's University Belfast, the Harvard-Smithsonian Center for Astrophysics, the Las Cumbres Observatory Global Telescope Network Incorporated, the National Central University of Taiwan, the Space Telescope Science Institute, the National Aeronautics and Space Administration under Grant No. NNX08AR22G issued through the Planetary Science Division of the NASA Science Mission Directorate, the National Science Foundation Grant No. AST–1238877, the University of Maryland, Eotvos Lorand University (ELTE), the Los Alamos National Laboratory, and the Gordon and Betty Moore Foundation.

\par This work has made use of data from the European Space Agency (ESA) mission {\it Gaia} (\url{https://www.cosmos.esa.int/gaia}), processed by the {\it Gaia} Data Processing and Analysis Consortium (DPAC, \url{https://www.cosmos.esa.int/web/gaia/dpac/consortium}). Funding for the DPAC has been provided by national institutions, in particular the institutions participating in the {\it Gaia} Multilateral Agreement.

\facility{Keck:II (DEIMOS), Keck:I (LRIS), Keck:I (HIRES), \Gaia,  Astro Data Lab, Magellan (MagE), Gemini North (GMOS)}

\software{\code{numpy} \citep{2011CSE....13b..22V,2020Natur.585..357H}, \code{matplotlib} \citep{2007CSE.....9...90H}, \code{scipy}  \citep{2020NatMe..17..261V}, \code{pandas} \citep{2022zndo...3509134T}}

\bibliography{main}
\let\clearpage\relax

\appendix
\onecolumngrid

\section{Membership Diagnostics for \unoshort{}}
\label{sec:appAdiagnostic}
In \figref{diagnostic} below, we present a four-panel summary of our DEIMOS spectroscopic catalogs covering \unoshort{}. The stellar radial velocities alone hold the most weight in defining the member sample. Note that this figure is the only place in this work for which we combined velocities across epochs.

\begin{figure*}[h]
    \centering
    \includegraphics[width=0.85\textwidth]{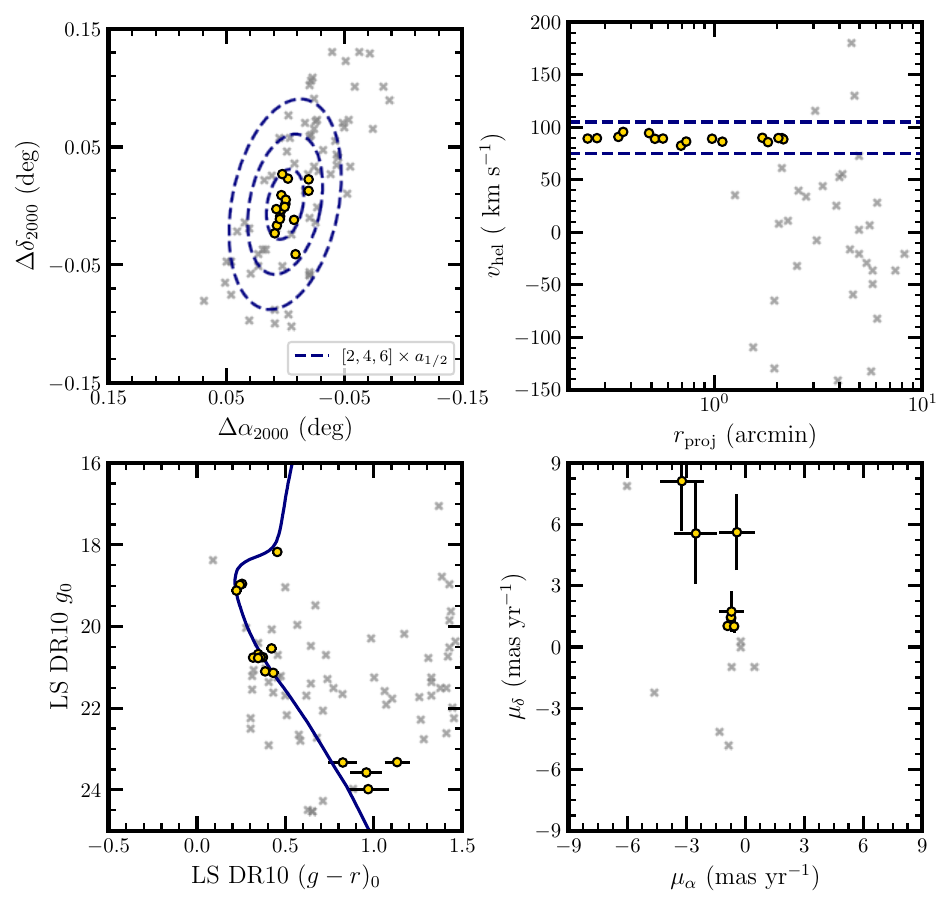}
    \caption{\textbf{Four diagnostic views of our complete, dual-epoch DEIMOS spectroscopic sample.} In each panel, \unoshort{} members are shown in gold while non-members are shown as grey crosses. (Top Left) Spatial positions of stars in a $0.3\degree \times 0.3\degree$ square region centered on \unoshort{}. Contours depicting $2,4,6\times$ the system's semi-major axis ($a_{1/2}$) are shown in blue. (Top Right) Velocity vs. projected radius for the same samples of stars.  (Bottom Left) Color--magnitude diagram based on Legacy Surveys DR10 photometry \citep{2019AJ....157..168D}, with a $\tau = 12$~Gyr, $\rm [Fe/H] = -2.19$ PARSEC isochrone overplotted in blue. We draw attention to the impressive depth of the DEIMOS observations -- reaching $g_0 \sim 24$. (Bottom Right) \Gaia DR3 proper motion vector-point diagram for the subset of brighter stars with measurements available. Although three stars have larger proper motion errors, there is a clear clustering among the remaining stars.  }
    \label{fig:diagnostic}
\end{figure*}

\newpage
\FloatBarrier

\newpage
\FloatBarrier
\section{Procedure for Deriving Frequentist Velocity Dispersion Upper Limits}
\label{sec:appBfrequentist}
As introduced in \secref{vdisp}, we derived upper limits on \unoshort{}'s velocity dispersion based on a frequentist profile likelihood approach. Here, we elaborate on the statistical procedure used to compute these limits. 
\par Following Section 2.4 of \citet{2011EPJC...71.1554C},  we computed the likelihood ratio test statistic
\[
q_{\sigma_v} =
\begin{cases}
-2 \ln \lambda(\sigma_v), & \text{if } \hat{\sigma}_v \le \sigma_v,\\[2mm]
0, & \text{if } \hat{\sigma}_v > \sigma_v.
\end{cases}
\]
over our logarithmic grid of $0.05~\kms \leq \sigma_v \leq 10~\kms$. This one-sided construction ensures that only dispersions larger than the best-fit value are considered when setting an upper limit.  In the large-$N$ limit and for an unbounded parameter $\sigma_v$, $q_{\sigma_v}$ asymptotically approach a $\chi^2$ distribution with one degree of freedom according to Wilks' theorem \citep{10.1214/aoms/1177732360}, i.e., $q_{\sigma_v} \rightarrow \chi^2_1$. In that case, the one-sided confidence upper limit could be found by simply looking up the critical value for the desired confidence level for a $\chi^2$ distribution; this is $q_{\sigma_v} = 2.71$ for a 95\% confidence level upper limit (one-sided). However, our dataset is smaller and physically bounded ($N=11-14$; $\sigma_v > 0.05$~\kms{}), and thus the Wilks theorem asymptotic behavior does not necessarily apply. 
\par Given this, we opted to empirically calibrate the  appropriate critical threshold for a 95\% confidence level upper limit with toy Monte Carlo simulations to ensure proper coverage.  For each trial value of $\sigma_v$ in our grid, we generated 3000 synthetic DEIMOS-like datasets by drawing from normal distributions $\mathcal{N}\left(\hat{v}(\theta\right), \sqrt{\sigma_v^2 + \epsilon_v^2})$ where $\hat{v}(\theta)$ is the maximum-likelihood of $v_{\rm sys}$ profiled over $\sigma_v$ from the real observed data, and $\epsilon_v$ were the real per-star observational errors for the Complete Epoch 2 ($N=14$) or Pure Epoch 2 ($N=11$) datasets. We then evaluated the test statistic for each simulated dataset ($q_{\sigma_v}^{\rm sim}$) and for the real observed data ($q_{\sigma_v}^{\rm obs}$) for each trial $\sigma_v$ value. Finally, we determined a formal 95\% confidence upper limit as the smallest dispersion for which the fraction of simulated datasets with $q_{\sigma_v}^{\rm sim} \geq q_{\sigma_v}^{\rm obs}$ -- i.e., the $p$-value -- fell below a threshold of 0.05.  In practice, this simulation-based approach yielded limits that were more conservative than if we simply had adopted the asymptotic limits, emphasizing the importance of the coverage calibration. More specifically, adopting the Wilks' theorem asymptotic condition yielded a limit of $\sigma_v < 1.8$~\kms{} for the Complete Epoch 2 sample, to be compared to our coverage-calibrated limit of $\sigma_v < 2.3$~\kms{}. For the Pure Epoch 2 sample, the Wilks' theorem condition was $\sigma_v < 1.4$~\kms{}, which can be compared to our coverage-calibrated limit of $\sigma_v < 1.6$~\kms{}.

\newpage
\FloatBarrier
\section{Ca II K Spectra for All LRIS Program Targets}
\label{sec:appCLRISspectra}
\begin{figure*}[htbp]
    \centering
    \includegraphics[width=0.65\textwidth]{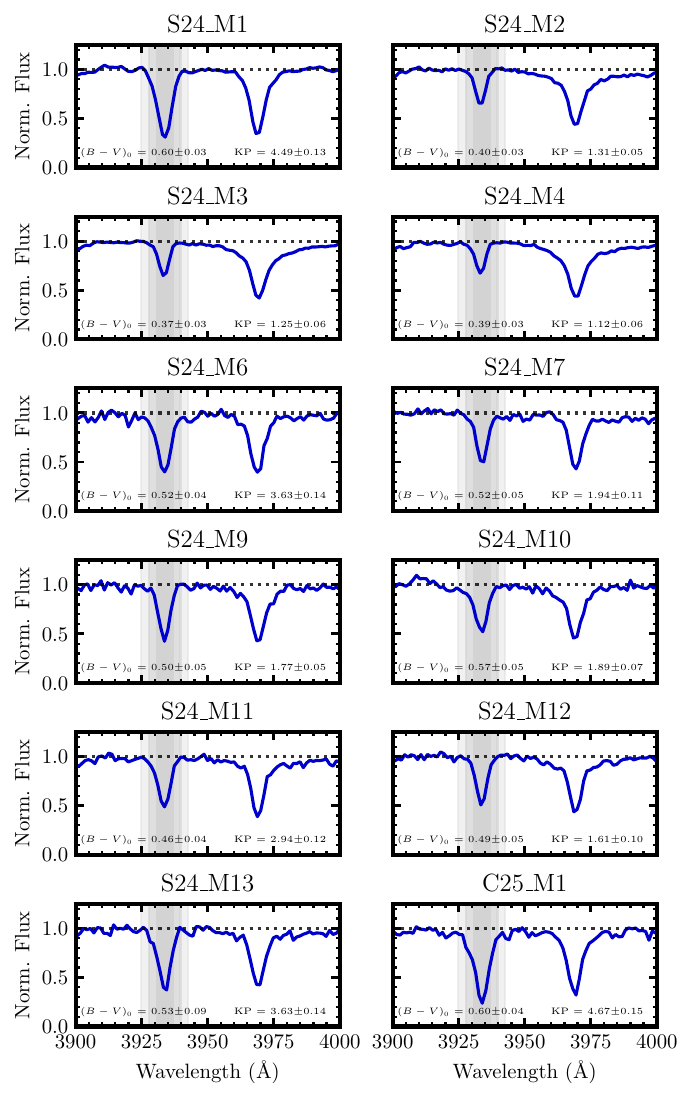}
    \caption{Normalized spectra for 12 stars observed with LRIS, limited to a wavelength range of $3900$--$4000\rm \AA$. In each panel, we shade the $[6,12,18] \ \rm \AA$ bands used for the KP index pseudo-EW measurements. In general, both our detailed qualitative inspection of these spectra and our EW measurements suggest there is no strong evidence for iron abundance dispersions. However, we note three stars (\texttt{S24\_M6}, \texttt{S24\_M11}, and \texttt{S24\_M13}) should be prioritized for future investigation. \texttt{S24\_M11}, in particular, has a noticeably high KP index than other stars at similar color, but this can be explained if it is a $2\sigma$ color outlier. We note this star is a candidate binary, and blending with a companion may also affect its measured EW at low spectral resolution and its color. }
    \label{fig:all_lris_spectra}
\end{figure*}

In \figref{all_lris_spectra}, we present the complete set of LRIS spectra used for our metallicity analysis.

\FloatBarrier
\section{Provenance of Measurements Presented in Figure 4}
\label{sec:appDcomparison}
In the bulleted list below, we summarize the original sources of the velocity dispersion and metallicity dispersion measurements presented in \figref{summary}. We stress that this is not a complete compilation of objects with $M_V > -3.5$; $r_{1/2} < 35$~pc, as objects with very weak limits, uncertain member samples, or no spectroscopic measurements have been excluded. The list below is ordered alphabetically. 

\begin{itemize}
     \item \textbf{BLISS 1}: We adopt the velocity dispersion limit of $\sigma_v < 5.5$~\kms{} (95\% credible level) from Cerny et al. 2025a, in prep. based on $N=5$ member stars observed with Magellan/IMACS. We also adopt the metallicity dispersion limit of from the same work, $\sigma_{\rm [Fe/H]} < 0.89$~dex based on a subset of just $N=3$ stars. The system is very likely a star cluster on the basis of its high metallicity (Cerny et al. 2025a, in prep.).
    \item \textbf{Bootes II}: We adopt the velocity dispersion of $\sigma_v = 2.9^{+1.6}_{-1.2}$~\kms{} based on 9 members observed with Magellan/IMACS by \citet{2023ApJ...950..167B}. From a subset of 8 of these stars, \citet{2023ApJ...950..167B} also reported a metallicity dispersion limit of $\sigma_{\rm [Fe/H]} < 0.47$~dex (95\% credible level). Bootes II is a confirmed dwarf galaxy; see also \citet{2014ApJ...794...89K,2016ApJ...817...41J,2025A&A...698A..63L}.
    \item \textbf{DELVE 1}: We adopt the velocity dispersion limit of $\sigma_v < 3.2$~\kms{} (95\% credible level) reported by \citet{2024ApJ...976..256S} from  $N = 8$ stars observed with Magellan/IMACS. This is in quantitative agreement with the limit of $\sigma_v < 3.9$~\kms{} from an overlapping 10-star Keck/DEIMOS dataset (Cerny et al. 2025a, in prep.). No metallicity dispersion has yet been published for the system, and its true classification remains unclear.
    \item \textbf{Draco II}: We adopt the velocity dispersion limit of $\sigma_v < 2.7$~\kms{} (95\% credible level) based on  $N = 27$ stars observed with Keck/DEIMOS (Cerny et al. 2025a, in prep.). We adopt the photometric metallicity dispersion of $\sigma_{\rm [Fe/H]} = 0.40\,\pm\,0.12$~dex from \citet{2023ApJ...958..167F}. Draco~II is likely a dwarf galaxy based on this latter measurement, but see Cerny et al. 2025a, in prep. for a more extended discussion.
    \item \textbf{ESO280-SC06}: We adopt the velocity dispersion of $\sigma_v = 2.27 \pm 0.38$~\kms{} from \citet{2019MNRAS.490..741S}, which was derived from 19 RGB members observed with AAT/AAOmega. We further adopt the metallicity dispersion upper limit of [Fe/H] $<0.11$ from Table 5 of \citet{usman2025chemical}, which was derived from $N=10$ stars observed at high resolution with Magellan/MIKE. ESO280-SC06 is a globular cluster, as is most clearly evident from its complete chemical abundance pattern \citep{usman2025chemical}.
    \item \textbf{Horologium~I}: We adopt both the velocity dispersion ($\sigma_v = 4.9^{+2.8}_{-0.9}$~\kms{}) and metallicity dispersion ($\sigma_{\rm [Fe/H]} = 0.17^{+0.20}_{-0.03}$~dex) estimates from \citet{2015ApJ...811...62K}, each based on $N=5$ stars.  These measurements demonstrate that Horologium~I is a confirmed ultra-faint dwarf galaxy (see also \citealt{2018ApJ...852...99N}).
    \item \textbf{Laevens 3}: We adopt the velocity dispersion limit of $\sigma_v < 4.45$~\kms{} (95\% credible level) reported by Cerny et al. 2025a, in prep, based on Keck/DEIMOS observations of 8 kinematic members. We adopt the joint photometric-spectroscopic metallicity dispersion limit of $\sigma_{\rm [Fe/H]} <0.5$  (95\% credible level) from  \citet{2019MNRAS.490.1498L}. Both studies concur Laevens~3 is most likely a star cluster.
    \item \textbf{Kim 3}:   We adopt the velocity dispersion limit of $\sigma_v < 3.0$~\kms{} (95\% credible level) from  $N = 7$ stars observed with Magellan/IMACS (Cerny et al. 2025a, in prep.).  No metallicity dispersion has yet been measured for the system, and its true classification remains unclear.
    \item \textbf{Palomar 13}: We adopt the velocity dispersion  limit of $\sigma_v < 1.45$~\kms{} (95\% credible level) from Geha et al., (2025; in press) based on a sample of $N= 99$ total members. We further adopt a metallicity dispersion limit of $\sigma_{\rm [Fe/H]} < 0.3$~dex based on the results of  \citet{2011ApJ...743..167B}; this is our own approximate estimate based on their reported $\sigma_{\rm [Fe/H]} = 0.1 \pm 0.1$~dex and one-sigma limit of $\sigma_{\rm [Fe/H]} < 0.2$~dex. Palomar 13 is a globular cluster.
    \item \textbf{Segue 1}: We adopt the velocity dispersion of $\sigma_v = 3.7^{+1.4}_{-1.1}$~\kms{} reported by \citet{2011ApJ...733...46S} based on $N=71$ stars observed with Keck/DEIMOS. We further adopt the photometric metallicity dispersion of $\sigma_{\rm [Fe/H]} = 0.65^{+0.26}_{-0.21}$~dex reported by \citet{2023ApJ...958..167F} from $N=12$ stars.
    \item \textbf{Segue 2}:  We adopt the velocity dispersion limit of $\sigma_v < 2.6$~\kms{} (95\% confidence upper limit) derived by \citet{2013ApJ...770...16K} based on $N=25$ member stars. We further 
    adopt the metallicity dispersion of $\sigma_{\rm [Fe/H]} = 0.43$~dex from the same work. Segue~2 is a confirmed dwarf galaxy.
    \item \textbf{Segue 3}:  We adopt the velocity dispersion limit of $\sigma_v < 2.3$~\kms{} (95\% credible level) reported by Cerny et al. in prep. from  $N$ = 20 stars within $3r_{1/2}$ observed with Keck/DEIMOS.   
    No metallicity dispersion has yet been measured for the system. Segue 3 is very likely a star cluster on the basis of its young age and high metallicity.
\item \textbf{Tucana III}: As introduced in \secref{classification}, Tucana~III is a particularly important point of comparison to \unoshort{} because it is the only ultra-faint, compact Milky Way satellite for which stronger velocity dispersion and metallicity dispersion limits exist. It is currently unclear whether Tucana~III is a dwarf galaxy or a star cluster, in part because it is clearly tidally disrupting \citep{DrlicaWagner2015}. The literature generally favors a dwarf galaxy albeit with substantial uncertainty. Evidence in favor of a dwarf galaxy include (1) its large size \citep{2017ApJ...838...11S}, (2) a tentative metallicity gradient from calcium triplet metallicities \citep{2018ApJ...866...22L}, (3) a metallicity difference from high-resolution spectroscopy of five stars \citep{2019ApJ...882..177M}, and (4) chemical signatures that may be more similar to a dwarf galaxy than a star cluster \citep{2017ApJ...838...44H,2019ApJ...882..177M}.

\par Each piece of evidence has significant caveats:
(1) Tucana~III's large size could be due to substantial tidal disruption, as Tucana~III has interacted both with the LMC and the Milky Way's bar \citep{Shipp_2021};
(2) the calcium triplet metallicity dispersion is not statistically significant;
(3) the high-resolution metallicity difference relies on the brightest star ($g=16.09$) analyzed in \citet{2017ApJ...838...44H} being 0.3 dex more metal-rich than four stars that are about 1 mag fainter in \citet{2019ApJ...882..177M}. However the metallicity difference relies on using spectroscopic stellar parameters for the brightest star, and \citet{2019ApJ...882..177M} note that adopting photometric stellar parameters no longer results in a significant metallicity difference. This is because the spectroscopic parameters assign nearly identical temperatures and gravities to the brightest star and the fainter stars, which cannot be correct if the stars are in the same stellar system.
(4) the detailed [X/Fe] abundance ratios are overall more similar to other metal-poor globular clusters than ultra-faint dwarf galaxies. In particular, the lack of multiple populations is similar to the low fraction of such stars in other globular cluster streams, which can be explained by an intrinsically low stellar mass \citep{2024MNRAS.529.2413U,usman2025chemical}. Additionally, the moderate r-process enhancement $\mbox{[Eu/Fe]}=+0.4$ is similar to that seen in other metal-poor globular clusters \citep{Kirby2023,usman2025chemical}, while nearly all ultra-faint dwarf galaxies have low neutron-capture element abundances \citep{2019ApJ...870...83J}.
While more studies are clearly needed, we suggest that a tidally disrupting stellar system where five stars have identical metallicities and chemical abundances is more likely to be a low mass star cluster. These conclusions corroborate the prior re-analysis by \citet{2025ApJ...987..217Z}.

\item \textbf{Tucana V}: We adopt the velocity dispersion of $\sigma_v= 1.2^{+0.9}_{-0.6}$~\kms{} from \citet{2024ApJ...968...21H}.  However, this dispersion is marginally resolved and an upper limit may be more appropriate. Recognizing this, \citet{2024ApJ...968...21H} quote an alternative measurement of an upper limit of $\sigma_v < 3.1$~\kms{} at the 95.5\% credible level. We  adopt the photometric metallicity of $0.43^{+0.44}_{-0.28}$ dex from \citet{2023ApJ...958..167F}. Tucana V is a confirmed dwarf galaxy based on its clear metallicity and abundance spreads (from both of the aformentioned works).

\item \textbf{Triangulum~II}: We adopt the velocity dispersion upper limit of 3.4~\kms{} (95\% credible level) from \citet{2022MNRAS.514.1706B} based on a sample of 16 stars observed with MMT/Hectochelle and Keck/DEIMOS (many with multi-epoch measurements). We adopt the metallicity dispersion of $\sigma_{\rm [Fe/H]} = 0.53^{+0.38}_{-0.12}$~dex from \citet{2017ApJ...838...83K}. Triangulum II is a confirmed dwarf galaxy on the basis of its metallicity spread (see also \citealt{2017MNRAS.466.3741V,2019ApJ...870...83J}).

\item \textbf{Willman~I}: We adopt the velocity dispersion of $\sigma_v =4.0 \pm0.8$~\kms{} derived by \citet{2011AJ....142..128W} based on 40 likely member stars (see their Section 5.1). We further adopt the photometric metallicity dispersion of $\sigma_{\rm [Fe/H]} = 0.65^{+0.10}_{-0.09}$ from 68 member stars reported by \citet{2023ApJ...958..167F}. Willman I is most likely a dwarf galaxy, albeit one with unusual kinematic and chemical properties. 
\end{itemize}

We acknowledge the use of the Local Volume Database \citep{Pace2024arXiv241107424P} for compiling many of the measurements discussed above.

\end{document}